\documentclass[journal]{IEEEtran}

\usepackage{graphicx}
\usepackage{subcaption} 
\usepackage{epstopdf}
\usepackage{dcolumn}
\usepackage{bm}
\usepackage{color}
\usepackage[mathlines]{lineno}
\usepackage[breaklinks=true,colorlinks=true,linkcolor=blue,urlcolor=blue,citecolor=blue]{hyperref}
\usepackage{ragged2e}
\usepackage{enumitem}
\usepackage{amsthm}
\usepackage{amssymb}
\usepackage{amsmath}
\usepackage{tabularx}
\usepackage{algorithm}
\usepackage{algpseudocode}

\newtheorem{thm}{Theorem}[section]

\theoremstyle{definition}

\theoremstyle{remark}

\newcolumntype{Y}{>{\centering\arraybackslash}X}

\ifCLASSINFOpdf
\newtheorem{thm}{Theorem} 
\else
\fi

\begin{document}


\title{Evolutionary Dynamics of Reputation-Based Voluntary Prisoner's Dilemma Games}

\author{Chen~Shen, Zhao~Song, Xinyu~Wang, Lei~Shi, Matja{\v z}~Perc, Zhen~Wang,~\IEEEmembership{Fellow,~IEEE}, and Jun~Tanimoto
\thanks{C. Shen and J. Tanimoto are with the Faculty of Engineering Sciences, Kyushu University, Fukuoka 816-8580, Japan. (email: steven\_shen91@hotmail.com; tanimoto.jun.380@m.kyushu-u.ac.jp).}
\thanks{Z. Song is with the Teesside University, Middlesbrough, UK. (email: z.song@tees.ac.uk).}
\thanks{X. Wang is with the CSSC Systems Engineering Research Institute and CSSC Intelligent Innovation Research Institute, Beijing, 100094, China (email: npu\_xinyu@163.com).}
\thanks{L. Shi is with the School of Statistics and Mathematics, Yunnan University of Finance and Economics, Kunming, Yunnan 650221, China. (Corresponding author; email: shi\_lei65@hotmail.com).}
\thanks{M. Perc is with the Faculty of Natural Sciences and Mathematics, University of Maribor, Maribor, Slovenia, and with the Community Healthcare Center Dr. Adolf Drolc Maribor, Maribor, Slovenia, and with the Department of Physics, Kyung Hee University, Seoul, Republic of Korea, and with the University College, Korea University, Republic of Korea (Corresponding author; email: matjaz.perc@gmail.com).}
\thanks{Z. Wang is with the School of Cybersecurity, and School of Artificial Intelligence, OPtics and ElectroNics (iOPEN), Northwestern Polytechnical University, China (e-mail: w-zhen@nwpu.edu.cn).}
}

\maketitle

\begin{abstract}
Cooperation underlies many natural and artificial systems. While voluntary participation can sustain cooperation without informational assumptions, real interactions are rarely anonymous, leaving the joint effects of participation and reputation insufficiently understood. We propose a reputation-based voluntary Prisoner’s Dilemma in which agents incur a monitoring cost to inspect opponents and decide whether to exit an interaction for a fixed incentive to avoid exploitation or to default to cooperation or defection. We show that reputation-conditioned exit generates multiple coexistence pathways that sustain cooperation across population structures. In well-mixed populations, cooperation persists through stable mixed coexistence, whereas in structured populations, exit-incentive–dependent regimes emerge, including local cyclic dominance and persistent oscillations. Together, these results extend voluntary participation frameworks and underscore the role of exit-incentive design in cooperative multi-agent systems.

\end{abstract}


\IEEEpeerreviewmaketitle
\section{Introduction}
Cooperation is fundamental to many complex systems, ranging from collective hunting in biological populations to global pandemic response in human societies and multi-agent coordination in artificial systems~\cite{fehr2002strong,sigmund2010social,chiong2012effects,chica2017networked}. In many such systems, agents face social dilemmas in which actions that maximize collective performance require individual sacrifice, whereas self-interested behavior yields immediate personal benefits. The Prisoner’s Dilemma provides a canonical abstraction of this conflict~\cite{nowak2006five,zhang2019strategy,li2016changing}: although mutual cooperation maximizes collective welfare, individually rational behavior favors defection, often leading to collectively suboptimal outcomes. Understanding how cooperative behavior can emerge and persist is therefore a central challenge in the design and control of complex systems.

Most established mechanisms for sustaining cooperation in evolutionary game theory rely on the availability of information that allows individuals to distinguish cooperators from defectors and condition their behavior accordingly~\cite{pepper2002mechanism,rand2013human}. Explicit incentive mechanisms, such as peer or institutional punishment~\cite{zhang2022peer,li2011game}, rewards~\cite{yang2017promoting,rand2009positive}, and social exclusion~\cite{sasaki2013evolution,liu2020evolutionary}, require identifying noncooperators to impose sanctions, but incur enforcement costs and are vulnerable to second-order free riding or corruption~\cite{lee2019social,herrmann2008antisocial}. Implicit mechanisms~\cite{nowak2006five}, including direct~\cite{van2012direct}, indirect~\cite{nowak2005evolution,gao2012cooperation,chong2007multiple}, and network reciprocity as well as group and kin selection~\cite{nowak1993spatial,olfati2007consensus,guo2023third,wang2009cooperative,platt2009kin}, similarly depend on information about identities, past actions, or interaction structure. Despite their differences, these approaches share a common reliance on informational prerequisites.

In contrast, voluntary participation has been shown to sustain cooperation under an information-free assumption~\cite{hauert2002volunteering,hauert2002replicator,zhang2023emergence,han2017evolution}, by allowing agents to withdraw from interactions without conditioning on opponents’ identities or behaviors. This exit option can generate cyclic dominance among cooperators, non-cooperators, and exiters, thereby preventing defection from becoming absorbing and allowing cooperation to persist. Subsequent studies have developed variants of voluntary participation~\cite{shen2021exit,arenas2011joker,gross2020self} and extended this mechanism across diverse settings, including collective-risk dilemmas~\cite{gross2019individual,mori2024outside}, costly punishment~\cite{hauert2007via,brandt2006punishing}, and repeated interactions~\cite{izquierdo2010option,izquierdo2014leave}. However, in models of one-shot interactions, exit is typically treated as unconditional, with participation decisions made independently of any prior information about opponents.

Viewed from a systems perspective, this modeling choice represents an idealized baseline rather than a realistic description of many engineered and social environments. In practice, interactions are rarely anonymous, and agents often possess at least coarse prior information about prospective partners through observation, reputation systems, or communication protocols~\cite{santos2018social,ren2023reputation,ren2025beyond}. For example, agents in distributed computing systems may decide whether to engage in task sharing based on peers’ past reliability~\cite{xiong2004peertrust,kamvar2003eigentrust}, while participants in online or human–AI collaborations may selectively enter interactions based on reputation or performance signals~\cite{liu2016simulation,pires2024artificial,Ashktorab2020Human}. In such environments, information shapes not only strategic behavior within interactions, but also participation decisions themselves. Reputation enables partner discrimination, whereas voluntary participation sustains cooperation through cyclic dominance; how conditioning exit on reputation alters evolutionary dynamics and cooperative outcomes remains largely unexplored.

To address this question, we propose a reputation-based one-shot voluntary Prisoner’s Dilemma that extends the standard voluntary Prisoner’s Dilemma by relaxing the assumption of indiscriminate exit and integrating reputation information into participation decisions. The population consists of four strategies: unconditional cooperators and defectors, and conditional cooperative and conditional defective exiters. Conditional exiters incur a monitoring cost to inspect opponents’ current strategies, interpreted as simple reputation signals, and decide whether to exit the interaction for a fixed incentive to avoid exploitation or to default to cooperation or defection.

Through replicator dynamics and agent-based modeling, we show that reputation-based exit admits a continuum of mixed equilibria involving all four strategies in well-mixed populations, provided that exit incentives exceed monitoring costs. In structured populations, including regular lattices, small-world networks, and random regular graphs, this requirement is relaxed, and reputation-based exit opens multiple exit-incentive–dependent coexistence pathways sustaining cooperation. These regimes involve cooperators, defectors, and both types of exiters, and include local cyclic dominance and oscillatory dynamics, with their emergence jointly mediated by exit incentives and population structure.

Our main contributions are summarized as follows:
\begin{itemize}
\item We propose a reputation-based one-shot voluntary Prisoner’s Dilemma model that incorporates reputation-conditioned exit decisions into the Prisoner’s Dilemma with voluntary participation.
\item We identify conditional exiters as key mediators of cooperative coexistence across population structures.
\item We demonstrate that exit incentives and population structure jointly determine cooperative coexistence patterns, ranging from stable coexistence to cyclic and oscillatory dynamics, with implications for cooperative multi-agent systems.
\end{itemize}

\section{Model}

\subsection{Voluntary Prisoner’s Dilemma Game with Reputation}
We propose a reputation-conditioned extension of the voluntary Prisoner’s Dilemma, building on the classical pairwise Prisoner’s Dilemma in which two players simultaneously choose to cooperate ($C$) or defect ($D$). Mutual cooperation yields a reward $R$, mutual defection yields a punishment payoff $P$, and mismatched choices result in a sucker’s payoff $S$ for the cooperator and a temptation payoff $T$ for the defector. We adopt the weak Prisoner’s Dilemma formulation~\cite{nowak1993spatial}, parameterized as $R = 1$, $P = S = 0$, and $T = b$, where $1 < b \leq 2$ represents the dilemma strength~\cite{wang2015universal}. This parametrization preserves the payoff ranking of the Prisoner’s Dilemma while allowing the incentive to defect to be tuned by a single parameter.

To introduce reputation-conditioned participation, players may, before each interaction, pay a monitoring cost $c$ to observe their opponent’s current strategy, which serves as a simple reputation signal. Based on this information, a player either plays the game or exits to avoid potential exploitation. If a player exits, the interaction is terminated: the exiting player receives a fixed, small payoff $\epsilon$ ($0 < \epsilon < 1$), interpreted as an exit incentive capturing capturing the reservation value of non-participation, while the non-exiting player receives zero payoff. This exit option captures unilateral abandonment of interactions, rather than continued participation without contribution~\cite{shen2021exit}, and is treated as behaviorally and strategically distinct from defection.

Introducing reputation-conditioned exit expands the strategy space from two to four strategies:

\begin{enumerate}
\item Unconditional Cooperator ($C$): cooperates unconditionally and does not pay the monitoring cost.
\item Unconditional Defector ($D$): defects unconditionally and does not pay the monitoring cost.
\item Conditional Cooperative Exiter ($CE$): pays the monitoring cost $c$ to inspect the opponent; exits if the opponent is identified as a defector ($D$ or $DE$), and otherwise cooperates.
\item Conditional Defective Exiter ($DE$): pays the monitoring cost $c$ to inspect the opponent; exits if the opponent is identified as a defector, and otherwise defects.
\end{enumerate}

Throughout this study, we restrict the monitoring cost to $0 < c < 1$, ensuring that monitoring and exit incur a nontrivial cost relative to the cooperative payoff and cannot trivially dominate unconditional strategies. The resulting interactions among the four strategies define the payoff matrix summarized in Table~\ref{t01}.

\begin{table}
\centering
\caption{\label{t01} Payoff matrix for the reputation-based voluntary Prisoner’s Dilemma. The extended model contains four competing strategies, which are unconditional cooperators ($C$) and defectors ($D$),as well as conditional cooperative exiters ($CE$) and conditional defective exiters ($DE$).
}
\begin{tabularx}{1.0\columnwidth}{l|YYYY} 
\hline
     & $C$ & $D$ & $CE$ & $DE$       \\
\hline
$C$  & 1 & 0 & 1 & 0\\
$D$  & $b$ & 0 & 0 & 0 \\
$CE$ & $1-c$    & $\epsilon-c$ & $1-c$ & $\epsilon-c$           \\
$DE$ & $b-c$ & $\epsilon-c$ & $-c$ & $\epsilon-c$ \\
\hline
\end{tabularx}
\end{table}

Importantly, the model is formulated as a one-shot interaction framework: agents access reputation information only at the moment of decision-making, and no memory of past encounters or repeated-interaction effects is assumed. Reputation is therefore operationalized as immediate strategy observability rather than as an accumulated or history-dependent attribute.

\subsection{Population structure}
We study four distinct population structures to evaluate the robustness of the proposed mechanism across different interaction topologies. A well-mixed population is represented by a complete graph, where each agent interacts with randomly chosen opponents from the entire population. The remaining three structures are degree-regular networks constructed from a square lattice with periodic boundary conditions, in which each agent has exactly four neighbors ($k=4$).

The regular square lattice serves as the baseline structured population. To interpolate between local and global interaction patterns while preserving degree regularity, we rewire its edges using the Watts–Strogatz algorithm~\cite{hauert2005game,watts1998collective}. Rewiring probabilities $p=0.01$ and $p=0.99$ generate small-world networks and random regular graphs, respectively. All network realizations preserve the same degree and differ primarily in characteristic path length, which is largest in the regular lattice, smallest in the random regular graph, and intermediate in the small-world network. Schematics of these networks are shown in Fig.~\ref{f02}A.

Evolutionary dynamics are governed by payoff-based selection, such that strategies with higher payoffs tend to increase in frequency, whereas those with lower payoffs decline. In infinite well-mixed populations, we analyze these dynamics using the replicator equation, which represents a strong-selection regime in which payoff differences directly determine changes in strategy frequencies. In networked populations, we study the same payoff-driven process through agent-based simulations, where agents update their strategies by imitating more successful neighbors. Detailed formulations of the replicator dynamics and simulation protocols are provided in subsequent sections.

\section{Infinite well-mixed populations}
\subsection{Replicator Dynamics}
We begin by examining the evolutionary dynamics in an infinite, well-mixed population. Let $x, y, z,$ and $w$ denote the population fractions of $C$, $D$, $CE$, and $DE$, respectively, which satisfy the constraint $x+y+z+w=1$. Under the assumption of random matching, the expected payoff of each strategy depends solely on the current population composition and the payoff matrix defined in Table~\ref{t01}. The expected payoffs are therefore given by:
\begin{equation}
\begin{aligned}
&\Pi_C  = x + z, \\
&\Pi_D  = bx, \\
&\Pi_{CE}  = x(1-c) + y(\epsilon-c) + z(1-c) + w(\epsilon-c), \\
&\Pi_{DE}  = x(b-c) + y(\epsilon-c) - zc + w(\epsilon-c).
\end{aligned}
\label{eq:payoffs}
\end{equation}

To describe the evolutionary dynamics, we employ replicator equations, under which strategy frequencies change according to their payoff advantage relative to the population average. Using the simplex constraint to eliminate $w$, the system reduces to the following three-dimensional dynamical system:
\begin{equation}
\begin{aligned}
&\dot{x} = x(\Pi_C - \overline{\Pi}):=f(x,y,z), \\
&\dot{y} = y(\Pi_D - \overline{\Pi}):=g(x,y,z), \\
&\dot{z} = z(\Pi_{CE} - \overline{\Pi}):=h(x,y,z)). \\
\end{aligned}
\label{eq:replicator}
\end{equation}
with $w=1-x-y-z$, and
$\overline{\Pi} = x\Pi_C + y\Pi_D + z\Pi_{CE} + w\Pi_{DE}$ denoting the average population payoff.

\subsection{Equilibrium and Stability Analysis}
By solving the replicator equation given by (\ref{eq:replicator}), we can derive several or equilibrium points.
\paragraph{Monomorphic equilibria} The four vertices of the simplex correspond to fixation of a single strategy:
\begin{itemize}
\item $x=1, y=0, z=0$, i.e., $E_1 = (1, 0, 0, 0)$, which means the dominance of $C$.
\item $x=0, y=1, z=0$, i.e., $E_2 = (0, 1, 0, 0)$, which means the dominance of $D$.
\item $x=0, y=0, z=1$, i.e., $E_3 = (0, 0, 1, 0)$, which means the dominance of $CE$).
\item $x=0, y=0, z=0$, i.e., $E_4 = (0, 0, 0, 1)$, which means the dominance of $DE$).
\end{itemize}

\paragraph{Polymorphic equilibria}
In addition, boundary and interior equilibria with strategy coexistence exist:
\begin{itemize}
\item $x=0, w=0$, i.e., $E_5 = \left( 0, \frac{c-1}{\epsilon-1}, \frac{\epsilon-c}{\epsilon-1}, 0 \right)$, representing coexistence of $D$ and $CE$.
\item $y=0, z=0$, i.e., $E_6 = \left( \frac{\epsilon-c}{\epsilon-b+1}, 0, 0, \frac{c-b+1}{\epsilon-b+1} \right)$, representing coexistence of $C$ and $DE$.
\item $E_7 = \left( \frac{\epsilon-c}{b\epsilon}, y^*, \frac{(\epsilon-c)(b-1)}{b\epsilon}, \frac{c}{\epsilon}-y* \right)$, with $y^*\in[0,\frac{c}{\epsilon}]$: a continuous manifold of interior equilibria involving all four strategies.
\end{itemize}
To determine local stability, we perform linear stability analysis using the Jacobian matrix of the reduced system with respect to $(x,y,z)$:
\begin{equation}
J = \begin{bmatrix}
\frac{\partial \dot f(x,y,z)}{\partial x} & \frac{\partial \dot f(x,y,z)}{\partial y} & \frac{\partial \dot f(x,y,z)}{\partial z} \\
\frac{\partial \dot g(x,y,z)}{\partial x} & \frac{\partial \dot g(x,y,z)}{\partial y} & \frac{\partial \dot g(x,y,z)}{\partial z} \\
\frac{\partial \dot h(x,y,z)}{\partial x} & \frac{\partial \dot h(x,y,z)}{\partial y} & \frac{\partial \dot h(x,y,z)}{\partial z}
\end{bmatrix}.
\label{eq:jacobian}
\end{equation}
An equilibrium point $E^*$ is locally asymptotic stable if all eigenvalues $\lambda_i$ of the Jacobian matrix satisfy $\text{Re}(\lambda_i) < 0$; it is unstable if at least one eigenvalue has a positive real part.  Non-hyperbolic equilibria are analyzed using center manifold theory.
Based on this analysis, we establish the following theorems.

\begin{thm}
    The equilibrium point $E_1=(1,0,0,0)$ is unstable.
\end{thm}
\begin{proof}
    The eigenvalues are $\{\lambda_1,\lambda_2,\lambda_3\}=\{b-1,b-c-1,-c\}$, where $\lambda_1>0$ as $b-1>0$. Thus $E_1$ is unstable.
\end{proof}

\begin{thm}
    The equilibrium point $E_2=(0,1,0,0)$ is locally asymptotically stable if and only if $\epsilon<c$.
\end{thm}
\begin{proof}
    The eigenvalues are $\left\{\lambda_{1}, \lambda_{2}, \lambda_{3}\right\}=\{0, \epsilon-c, \epsilon-c\}$. When $\epsilon>c$, and $E_2$ is unstable since $\lambda_2,\lambda_3>0$. When $\epsilon<c$, the equilibrium is non-hyperbolic due to the zero eigenvalue. We therefore apply the center manifold theorem. First, we shift the equilibrium point $E_2$ to the origin by defining the deviation variables $\mathbf{x}_1 = [x_1, y_1, z_1]^T$:
\begin{equation}
\left\{\begin{array}{l}x_{1}=x \\ y_{1}=y-1 \\ z_{1}=z\end{array}\right.
\end{equation}
We construct a transformation matrix $T$ whose columns are the eigenvectors of $J$ (where the first column corresponds to the zero eigenvalue), and its inverse $P = T^{-1}$:
\begin{equation}
T=\left[\begin{array}{ccc}-1 & 0 & 0 \\ 1 & 1 & 0 \\ 0 & 0 & 1\end{array}\right], \quad P=\left[\begin{array}{ccc}-1 & 0 & 0 \\ 1 & 1 & 0 \\ 0 & 0 & 1\end{array}\right]
\end{equation}
$P$ diagonalizes $J$ as:
\begin{equation}
J_{1} = P J P^{-1}=\left[\begin{array}{ccc}0 & 0 & 0 \\ 0 & \epsilon-c & 0 \\ 0 & 0 & \epsilon-c\end{array}\right]
\end{equation}

We introduce the new coordinate system $\mathbf{x}_2 = [x_2, y_2, z_2]^T$ defined by the linear transformation $\mathbf{x}_2 = P \mathbf{x}_1$. Substituting the shifted variables into this transformation yields:
\begin{equation}
\left[\begin{array}{l}x_{2} \\ y_{2} \\ z_{2}\end{array}\right]=P\left[\begin{array}{l}x \\ y-1 \\ z\end{array}\right]=\left[\begin{array}{c}-x \\ x+y-1 \\ z\end{array}\right]
\end{equation}

The dynamics in the new coordinates are given by $\dot{\mathbf{x}}_2 = P \dot{\mathbf{x}}_1$. Calculating the derivatives explicitly implies $\dot{x}_2 = -\dot{x}$, $\dot{y}_2 = \dot{x} + \dot{y} -1$, and $\dot{z}_2 = \dot{z}$. We separate the system into the Center part ($\boldsymbol{X}$) and the Stable part ($\boldsymbol{Y}$):
\begin{equation}\left\{\begin{array}{l}\dot{\boldsymbol{X}}=A \boldsymbol{X}+\boldsymbol{F}(\boldsymbol{X}, \boldsymbol{Y}) \\
\dot{\boldsymbol{Y}}=B \boldsymbol{Y}+\boldsymbol{G}(\boldsymbol{X}, \boldsymbol{Y})\end{array}\right.
\end{equation}
Where $\boldsymbol{X} = [x_2] \in \mathbb{R}^1$, $\boldsymbol{Y} = [y_2, z_2]^T \in \mathbb{R}^2$. The linear matrices are $A=[0]$ and $B=\text{diag}(\epsilon-c, \epsilon-c)$.

To determine stability, we need the explicit Taylor expansion of the nonlinear term $\boldsymbol{F}(\boldsymbol{X}, \boldsymbol{Y})$ corresponding to $\dot{x}_2$. From the replicator equation $\dot{x} = x(\Pi_C - \bar{\Pi})$, we approximate the dynamics on the center manifold by setting the stable variables to zero ($\boldsymbol{Y}=\mathbf{0}$).
From the coordinate mapping, $y_2=0$ implies $y=1-x$, and $z_2=0$ implies $z=0$, then the replicator equation for $x$ becomes:
\begin{equation}
  \dot{x} = x [ x - (x^2 + bx(1-x)) ] = (1-b)x^2 + (b-1)x^3.
\end{equation}
Using the relationship $x = -x_2$ derived from the transformation, the dynamics of the center variable $x_2$ are:
\begin{equation}
\dot{x}_2 = -\dot{x} \approx - [ (1-b)(-x_2)^2 ] = (b-1)x_2^2.
\end{equation}
So, the function $\boldsymbol{F}(\boldsymbol{X}, \boldsymbol{Y})$ restricted to the center manifold is:
\begin{equation}
\boldsymbol{F}(x_2, \mathbf{0}) = (b-1)x_2^2 + \mathcal{O}(x_2^3).
\end{equation}
Consider a small perturbation from the equilibrium $x_2 = 0$. Since $x_2 = -x$ and $x \ge 0$, the physical domain corresponds to $x_2 \le 0$. Let $x_2 = -\delta$ where $\delta > 0$. 
\begin{equation}
    \dot{x}_2 = (b-1) (-\delta)^2 = (b-1)\delta^2.
\end{equation}
Given $1 < b \le 2$, we have $(b-1) > 0$, and thus $\dot{x}_2 > 0$. Since $x_2$ is negative and its derivative is positive, $x_2$ increases towards $0$. This implies that the trajectory converges to the equilibrium restricted to the Center Manifold. Combined with the exponential decay in the stable directions ($\lambda_2, \lambda_3 < 0$), the equilibrium point $E_2$ is locally asymptotically stable.
\end{proof}

\begin{thm}
    The equilibrium point $E_3=(0,0,1,0)$ is unstable.
\end{thm}
\begin{proof}
    The eigenvalues are $\{\lambda_1,\lambda_2,\lambda_3\}=\{-1,c-1,c\}$, where $c>0$. Thus $E_3$ is unstable since $\lambda_3>0$.
\end{proof}

\begin{thm}
    The equilibrium point $E_4=(0,0,0,1)$ is unstable.
\end{thm}
\begin{proof}
The eigenvalues are $\left\{\lambda_{1}, \lambda_{2}, \lambda_{3}\right\}=\{c-\epsilon, c-\epsilon, 0\}$.
   When $\epsilon<c$, we have $\lambda_1, \lambda_2 > 0$, then $E_4$ is unstable. When $\epsilon>c$, we have $\lambda_1, \lambda_2 < 0$ while $\lambda_3 = 0$. Therefore, we determine the local stability through the centre manifold theorem.
First, we construct the transformation matrix $T$ whose columns are the eigenvectors of $J$, and its inverse $P$:
\begin{equation}
T=\left[\begin{array}{ccc}1 & 0 & 0 \\ 0 & 1 & 0 \\ 0 & 0 & 1\end{array}\right], \quad P=T^{-1}=\left[\begin{array}{ccc}1 & 0 & 0 \\ 0 & 1 & 0 \\ 0 & 0 & 1\end{array}\right]
\end{equation}
$P$ leaves $J$ unchanged:
\begin{equation}
J_{1} = P J P^{-1}=\left[\begin{array}{ccc}c-\epsilon & 0 & 0 \\ 0 & c-\epsilon & 0 \\ 0 & 0 & 0\end{array}\right]
\end{equation}

We introduce the new coordinate system $\mathbf{x}_2 = [x_2, y_2, z_2]^T$ defined by $\mathbf{x}_2 = P \mathbf{x}_1$. Since $P$ is the identity matrix, the coordinates remain unchanged:
\begin{equation}
\left[\begin{array}{l}x_{2} \\ y_{2} \\ z_{2}\end{array}\right] = \left[\begin{array}{l}x \\ y \\ z\end{array}\right]
\end{equation}

The dynamics are given by $\dot{\mathbf{x}}_2 = P \dot{\mathbf{x}} = \dot{\mathbf{x}}$. We separate the system into the Center part ($\boldsymbol{X}$) and the Stable part ($\boldsymbol{Y}$) based on the eigenvalues. Here, the third component corresponds to the zero eigenvalue:
\begin{equation}\left\{\begin{array}{l}\dot{\boldsymbol{Y}}=B \boldsymbol{Y}+\boldsymbol{G}(\boldsymbol{X}, \boldsymbol{Y}) \\
\dot{\boldsymbol{X}}=A \boldsymbol{X}+\boldsymbol{F}(\boldsymbol{X}, \boldsymbol{Y})\end{array}\right.
\end{equation}
Where $\boldsymbol{X} = [z_2] \in \mathbb{R}^1$ represents the center variable, and $\boldsymbol{Y} = [x_2, y_2]^T \in \mathbb{R}^2$ represents the stable variables. The linear matrices are $A=[0]$ and $B=\text{diag}(c-\epsilon, c-\epsilon)$.

\begin{figure*}[!t]
	\centering
	\includegraphics[width=0.91\textwidth]{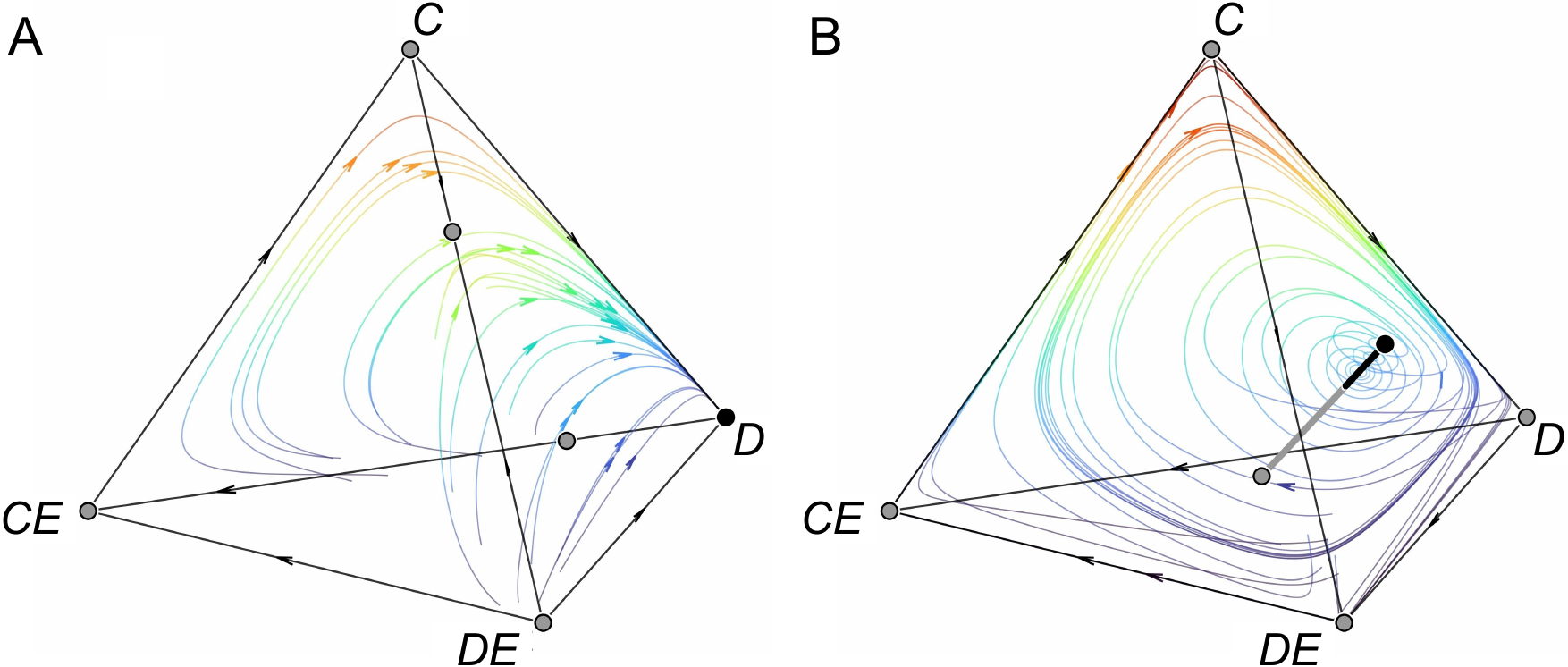}
	\caption{\textbf{Reputation-based exit sustains cooperation through stable coexistence when exit incentives exceed monitoring costs.} Phase portraits show four-strategy replicator dynamics on a two-dimensional simplex projection. (A) When monitoring costs exceed exit benefits ($c > \epsilon$), defection is the unique asymptotically stable equilibrium (black filled circle). (B) When exit benefits exceed monitoring costs ($c < \epsilon$), defection loses global stability and trajectories converge to a one-dimensional manifold of mixed equilibria $E_7$ (black line) involving cooperation, defection, and exit strategies ($CE$ and $DE$). Black filled circles denote representative stable equilibria on this manifold, whereas gray circles indicate unstable equilibria. Streamlines indicate the direction of selection, and color encodes evolutionary speed, with warmer colors indicating faster dynamics. Parameters are fixed at $b=1.5$ and $c=0.4$; the exit incentive is $\epsilon=0.2$ in (A) and $\epsilon=0.6$ in (B).
	}
	\label{fig:both_infinite}
\end{figure*}

To determine stability, we need the explicit Taylor expansion of the nonlinear term $\boldsymbol{F}(\boldsymbol{X}, \boldsymbol{Y})$ corresponding to $\dot{z}_2$. From the replicator equation $\dot{z} = z(\Pi_{CE} - \bar{\Pi})$, we approximate the dynamics on the center manifold by setting the stable variables to zero ($\boldsymbol{Y}=\mathbf{0}$).
Then the replicator dynamic equation becomes:
\begin{equation}
  \dot{z}_2 = \dot{z} \approx z ( \Pi_{CE} - \Pi_{DE} ) = z_2^2.
\end{equation}
So, the function $\boldsymbol{F}(\boldsymbol{X}, \boldsymbol{Y})$ restricted to the center manifold is:
\begin{equation}
\boldsymbol{F}(z_2, \mathbf{0}) = z_2^2 + \mathcal{O}(z_2^3).
\end{equation}

Consider a small perturbation from the equilibrium $z_2 = 0$. Since $z_2$ corresponds to the population density $z$, we require $z_2 \ge 0$. Let $z_2 = \delta$ where $\delta > 0$. 
\begin{equation}
    \dot{z}_2 = \delta^2 > 0.
\end{equation}
Since $z_2$ is positive and its derivative $\dot{z}_2$ is positive, $z_2$ will increase, driving the system away from the origin. The trajectory diverges to the equilibrium restricted to the Center Manifold. Therefore, the equilibrium point $E_4$ is locally unstable.
\end{proof}

\begin{thm}
    The equilibrium point $E_5=(0,\frac{c-1}{\epsilon-1},\frac{\epsilon-c}{\epsilon-1},0)$ is unstable.
\end{thm}
\begin{proof}
 $E_5$ exists when $\epsilon < c < 1$.
    Its eigenvalues are $\left\{\lambda_{1}, \lambda_{2}, \lambda_{3}\right\}=\left\{\frac{c-\epsilon}{\epsilon-1},\frac{\epsilon-c}{\epsilon-1},\frac{(c-1) (c-e)}{\epsilon-1}\right\}$, where $\frac{c-\epsilon}{\epsilon-1}$ and $\frac{\epsilon-c}{\epsilon-1}$ always have the opposite signs. Thus, $E_5$ is untable.
\end{proof}

\begin{thm}
    The equilibrium point $E_6=(\frac{\epsilon-c}{1-b+\epsilon},0,0,\frac{1-b+c}{1-b+\epsilon})$ is unstable.
\end{thm}
\begin{proof}
$E_6$ exists when $c<\epsilon$ and $b<1+c$, or $c>\epsilon$ and $b>1+c$
    The eigenvalues are $\left\{\frac{(b-1) (c-\epsilon)}{b-\epsilon-1},\frac{(b-1) (\epsilon-c)}{b-\epsilon-1},\frac{(-b+c+1) (c-\epsilon)}{b-\epsilon-1}\right\}$, where $\lambda_1$ and $\lambda_2$ always have the opposite signs. Thus, $E_6$ is unstable.
\end{proof}
\begin{thm}
    The equilibrium point $E_7 = \left( \frac{\epsilon-c}{b\epsilon}, y^*, \frac{(\epsilon-c)(b-1)}{b\epsilon}, \frac{c}{\epsilon}-y^* \right)$, $y^*\in[0,\frac{c}{\epsilon}]$ is unstable when $y^*\in[0,\frac{c}{b\epsilon})$, and stable when $y^*\in[\frac{c}{b\epsilon},\frac{c}{\epsilon}]$. At $y^*=\frac{c}{b\epsilon}$, $E_7$ is neutrally stable.
\end{thm}

\begin{proof}
    $E_7$ exists when $c<\epsilon$. Its eigenvalues are $\lambda_1=0$, and 
\begin{equation}
    \lambda_{2,3}(y^*) = \frac{\epsilon-c}{2b\epsilon} \left[ \underbrace{(c - b \epsilon y^*)}_{A(y^*)} \pm \sqrt{\underbrace{(c - b \epsilon y^*)^2 - \Omega}_{\Delta(y^*)}} \right]
\end{equation}
where $A(y^*)$ is the linear trace term, and $\Omega = 4bc(b-1) > 0$ is a positive constant as since $b>1, c>0$.
We analyze the sign of the real part of $\lambda_{2,3}$ in three cases: 

i) Case 1: complex eigenvalues ($A(y^*)\neq 0$, $\Delta(y^*) < 0$).
    This occurs when $A(y^*)^2 < \Omega$. The eigenvalues are complex conjugates: $\lambda = \frac{x^*}{2}(A \pm i\sqrt{|\Delta|})$.
    The stability is determined solely by the sign of $A(y^*)$, namely, $E_7$ is stable when $A(y^*) < 0$, otherwise, is unstable.
    
ii) Case 2: real eigenvalues ($A(y^*)\neq 0$, $\Delta(y^*) > 0$).
    This occurs when $A(y^*)^2 \ge \Omega$. The eigenvalues are real.
    Crucially, observe that since $\Omega > 0$, we have:
    \begin{equation}
        \sqrt{A(y^*)^2 - \Omega} < \sqrt{A(y^*)^2} = |A(y^*)|
    \end{equation}
    This inequality implies that the magnitude of the square root term is strictly less than the magnitude of the leading term $A(y^*)$. Therefore, the sign of both eigenvalues $\lambda_2$ and $\lambda_3$ is entirely determined by the sign of $A(y^*)$, because the $\pm$ term is not large enough to flip the sign. Therefore, $E_7$ is stable if $A(y^*) < 0$, otherwise, is unstable.
    
iii) Case 3: imaginary eigenvalues ($A(y^*)=0$, $\Delta(y^*) < 0$).
    This occurs when $y^* = \frac{c}{b\epsilon}$. The eigenvalues are complex conjugates: $\lambda = \frac{x^*}{2}(\pm i\sqrt{-\Delta})$.
    Therefore, $E_7$ is nuetral stable when $A(y^*) = 0$.

\end{proof}

\subsection{Summary of cooperation dynamics in well-mixed Populations}
Taken together, the replicator analysis demonstrates that reputation-based exit sustains cooperation only through stable coexistence when exit incentives exceed monitoring costs. When monitoring costs exceed exit incentives ($c>\epsilon$; Fig.~\ref{fig:both_infinite}A), unconditional defection is the unique asymptotically stable equilibrium, and all trajectories converge to it; in this regime, reputation-conditioned exit cannot offset its informational cost. By contrast, when the exit incentive exceeds the monitoring cost ($c<\epsilon$; Fig.~\ref{fig:both_infinite}B), defection loses global stability, and the dynamics converge to a one-dimensional manifold of mixed equilibria involving cooperation, defection, and exit strategies. The final population composition depends continuously on initial conditions, reflecting convergence along an equilibrium manifold rather than discrete multistability.



\section{Networked populations}
We now consider networked populations to examine how interaction topology shapes evolutionary dynamics beyond the well-mixed limit.
\subsection{Agent-based simulation}
We consider a network $G (V, E)$ of size $N$, where $V$ is the set of nodes and $E$ the set of links. Each node represents an agent, and edges define pairwise interaction and social-learning channels. An agent $i$ occupies a node and interacts only with its local neighborhood $\mathcal{N}_i$, as determined by network connectivity. Unlike the global and random interactions in well-mixed populations, networked interactions are local and structured, thereby limiting both interaction and social learning and embedding the evolutionary dynamics within a specific spatial topology.

\begin{figure*}[!htbp]
\centering
\includegraphics[width=1.7\columnwidth]{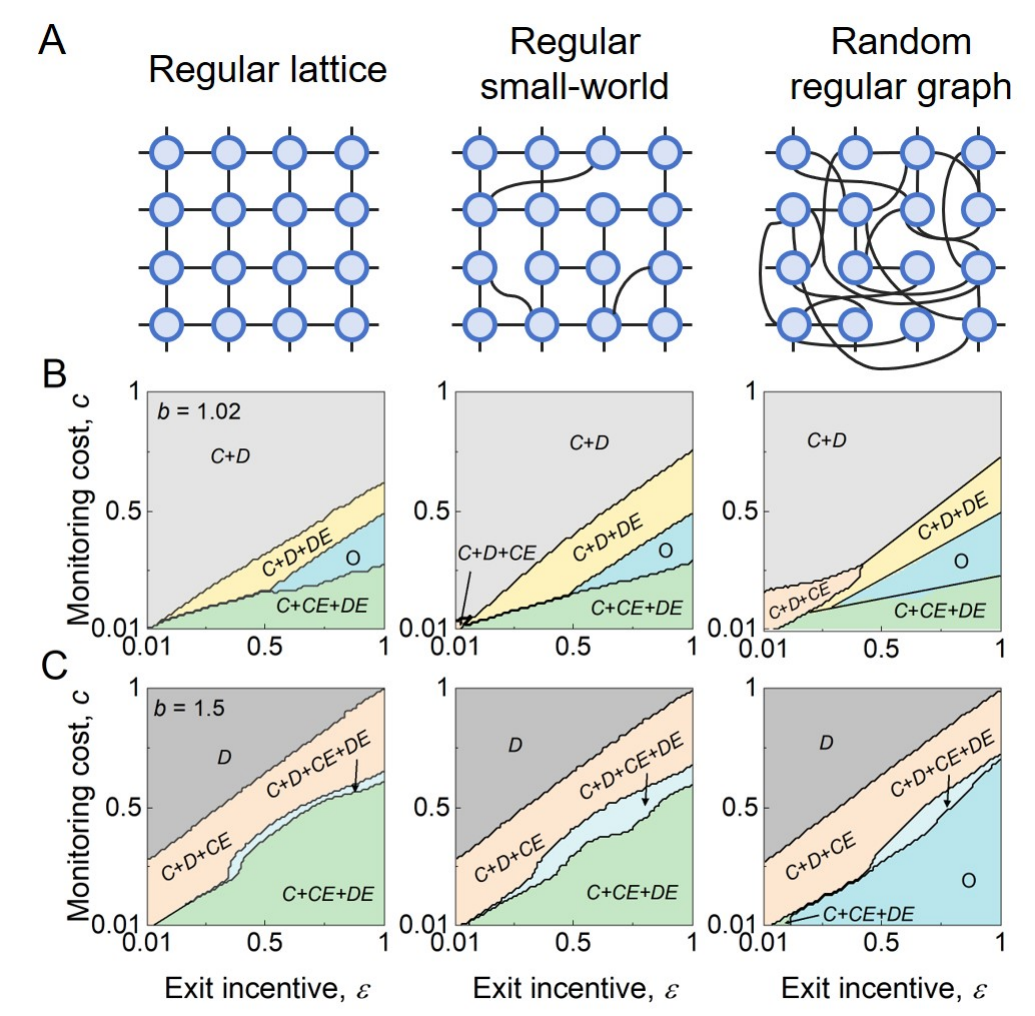}
\caption{\textbf{Structured populations enable the coexistence of cooperation and reputation-based exit strategies through multiple pathways.} (A) Schematic of the three degree-regular network topologies: regular lattice, regular small-world network, and random regular graph.
(B, C) Phase diagrams showing the stationary composition of strategies as functions of the exit incentive $\epsilon$ and monitoring cost $c$ for the regular lattice (left), regular small-world network (middle), and random regular graph (right). Results are shown for (B) a weak dilemma ($b = 1.02$) and (C) a strong dilemma ($b = 1.5$). Each labeled region (e.g., $C+D+CE$ or $C+CE+DE$) indicates the subset of strategies that persist with nonzero stationary frequency. Regions labeled $O$ denote parameter regimes exhibiting persistent oscillations with coexistence of all four strategies.}
\label{f02}
\end{figure*}

\begin{figure*}[!h]
\centering
\includegraphics[scale=0.7]{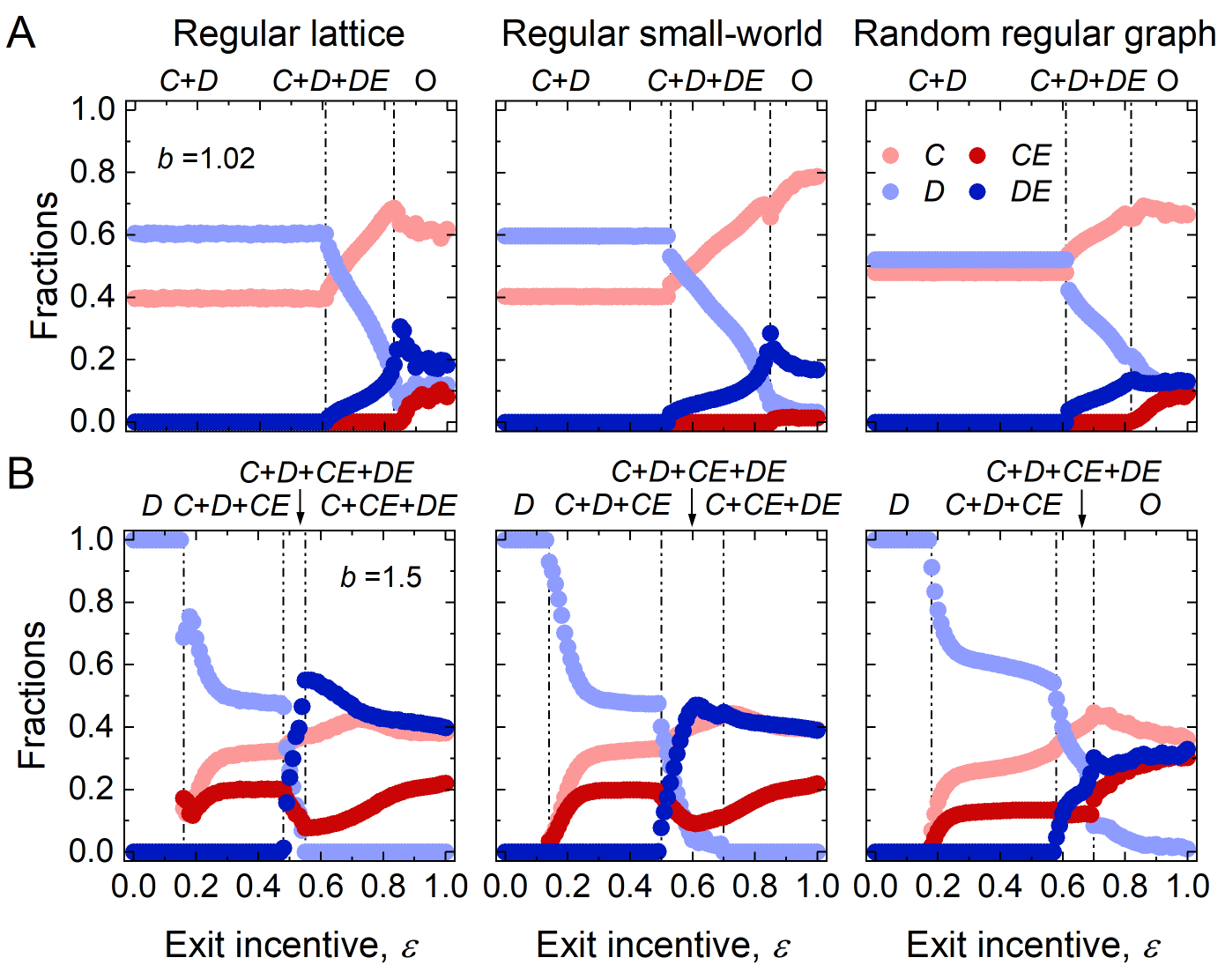}
\caption{\textbf{Adjusting exit incentives reshapes dominance among cooperators, defectors, and exit strategies in structured populations.} 
Shown are the stationary fractions of strategies as functions of the exit incentive $\epsilon$ for the regular lattice (left), regular small-world network (middle), and random regular graph (right).
Results are shown for (A) weak dilemma strength ($b=1.02$) and (B) strong dilemma strength ($b=1.5$). Vertical dashed lines indicate transitions between coexistence regimes identified in Fig.~2 and labeled above each panel. The monitoring cost is fixed at $c=0.4$ in all panels. Light and dark red denote $C$ and $CE$, respectively, while light and dark blue denote $D$ and $DE$.
}
\label{f03}
\end{figure*}

\begin{figure*}[!h]
    \centering
\includegraphics[width=0.8\textwidth]{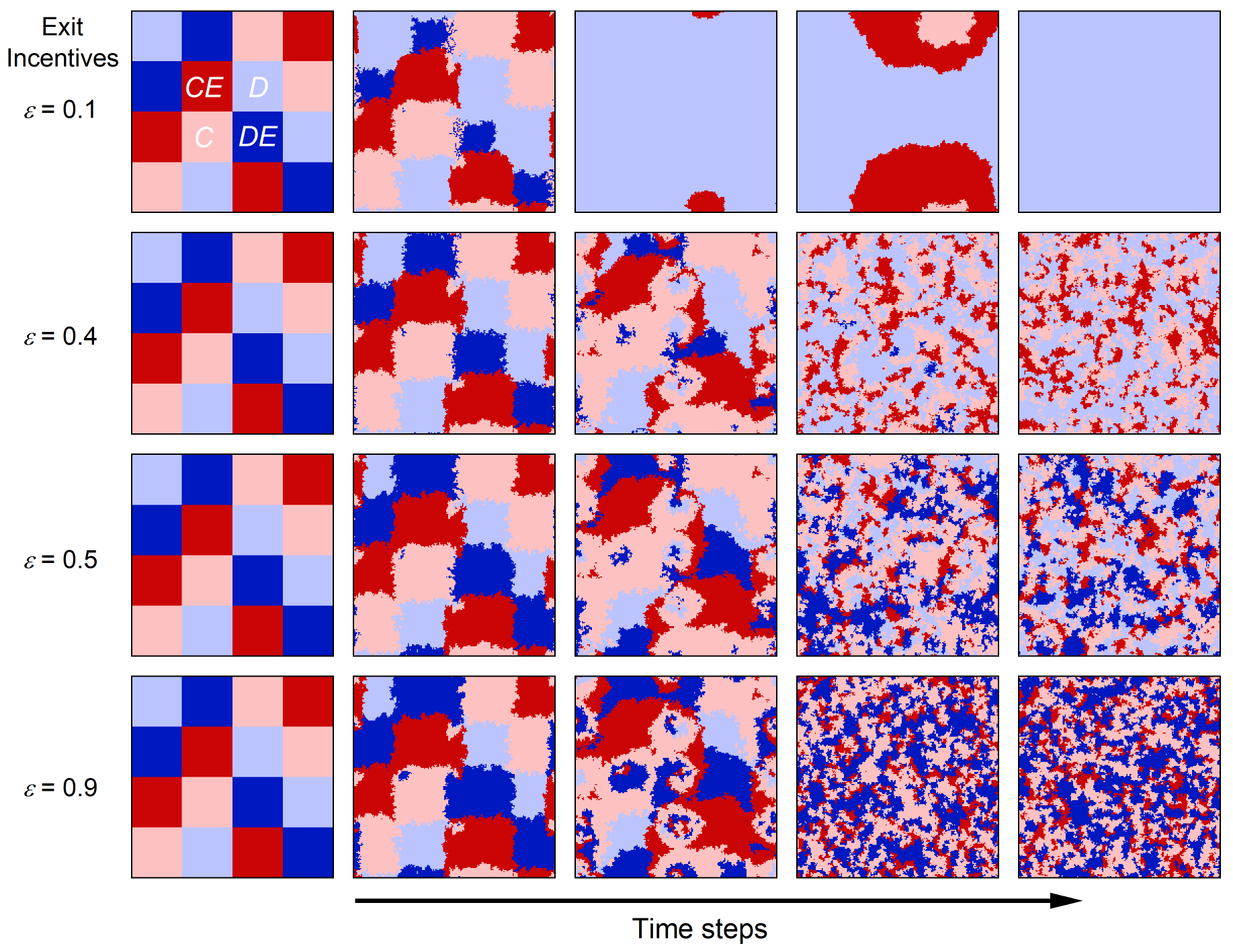}
\caption{\textbf{Conditional exiters mediate exit-incentive–dependent coexistence routes via local cyclic dominance.} 
Representative snapshots of spatial strategy configurations for exit incentives $\epsilon = 0.1, 0.4, 0.5$, and 0.9 (top to bottom). For each row, the first and last panels show the initial and final configurations, while the intermediate panels (second to fourth columns) correspond to selected intermediate time steps chosen for illustration. Simulations are performed on a square lattice with dilemma strength fixed at $b=1.5$ and monitoring cost at $c=0.4$. Colors denote strategies: $C$, light red; $D$, light blue; $CE$, red; $DE$, dark blue.
}
\label{f04}
\end{figure*}

For networked populations, the evolutionary process is implemented through agent-based simulations that explicitly resolve local interactions and imitation dynamics. In each elementary update, a focal player $i$ interacts with all neighbors in $\mathcal{N}_i$ to accumulate a total payoff $\Pi_i$. A randomly chosen neighbor $j \in \mathcal{N}_i$ accumulates its payoff $\Pi_j$ analogously. Player $i$ then updates its strategy by imitating player $j$ with probability:
\begin{equation}
W_{i\leftarrow{}j}=\frac{1}{1+\exp\left(\left(\Pi_i-\Pi_j\right)/\kappa\right)},
\label{eq-fermi}
\end{equation}
such that strategies with higher payoffs are more likely to be adopted. Here $\kappa \ge 0$ controls selection strength~\cite{sigmund2010social}: smaller values of $\kappa$ correspond to stronger selection, whereas larger $\kappa$ increases stochasticity. We set $\kappa=0.1$, corresponding to a strong-selection regime, to ensure comparability with well-mixed results.

Random exploration is introduced at rate $\mu=10^{-6}$ to generate behavioral variation and mitigate finite-size effects~\cite{shen2025mutation}. With probability $\mu$, agent $i$ adopts a randomly chosen strategy from $\{C,D,CE,DE\}$; otherwise, updating follows Eq.~\eqref{eq-fermi}. The full agent-based simulation procedure is summarized in Algorithm~\ref{simu-1}.

\begin{algorithm}[H]
\centering
    \caption{Agent-based simulation of the proposed reputation-based Voluntary Prisoner’s Dilemma}
    \label{simu-1}
    \small
    \algnewcommand\Input{\item[\textbf{Input:}]}
    \algnewcommand\Output{\item[\textbf{Output:}]}
    \begin{algorithmic}[1]
        \Input Network topology $G(V, E)$, Payoff matrix Table~\ref{t01}, Total time steps $T_{max}$, Averaging window $T_{avg}$.
        \Output Average fraction of cooperators $\rho_C$.
        
        \State \textbf{Initialization:} 
        Arrange players on nodes, and assign each player a random strategy.
        
        \For{$t = 1$ to $T_{max}$}
            \For{$i = 1$ to $N$} \Comment{One Monte Carlo Step (MCS)}
                \State Randomly select a focal player $i$ from $V$;
                \State Calculate total payoff $\Pi_i$ by interacting with  $\mathcal{N}_i$;
                \State Randomly select a neighbor $j$ from $\mathcal{N}_i$;
                \State Calculate total payoff $\Pi_j$ by interacting with $\mathcal{N}_j$;
                
                \State Generate a random number $r_1 \in [0, 1]$;
                \If{$r_1 < \mu$} \Comment{Exploration mechanism}
                    \State $s_i \leftarrow$ Random strategy from $\{C, D, CE, DE\}$
                \Else \Comment{Imitation mechanism}
                    \State Calculate Fermi probability:
                    \State $W \leftarrow \frac{1}{1+\exp\left[(\Pi_i - \Pi_j)/\kappa\right]}$;
                    \State Generate a random number $r_2 \in [0, 1]$;
                    \If{$r_2 < W$}
                        \State $s_i \leftarrow s_j$;
                    \EndIf
                \EndIf
            \EndFor
            
            \If{$t > T_{max} - T_{avg}$}
                \State Record current fraction of cooperators $\rho_C(t)$;
            \EndIf
        \EndFor
        
        \State Calculate average: $\rho_C \leftarrow \frac{1}{T_{avg}} \sum_{t=T_{max}-T_{avg}+1}^{T_{max}} \rho_C(t)$;
        \State \Return $\rho_C$
    \end{algorithmic}
\end{algorithm}

The simulation protocol is configured to ensure convergence to stationary states without systematic trends in strategy frequencies. A complete Monte Carlo step (MCS) comprises $N$ elementary updates, ensuring that every agent updates their strategy once on average. Simulations run for a sufficiently long duration to bypass transient dynamics, typically $50,000$ MCS, with final strategy abundances averaged over the last $5,000$ steps. Each data point is obtained by averaging over 10 independent realizations, and the population size is fixed at $N = 4 \times 10^{4}$ for all network structures.

\subsection{Evolutionary outcomes}
Structured populations can sustain cooperation via network reciprocity, whereby local interactions allow cooperators to cluster when the dilemma strength $b$ lies below a network-dependent critical value~\cite{nowak1993spatial}. We therefore focus on two representative cases, $b=1.02$ and $b=1.5$, corresponding to weak and strong dilemma strengths, respectively.

Relative to well-mixed populations, structured interactions generate multiple, topology-dependent coexistence pathways for cooperation and reputation-based exit. For the weak dilemma case ($b=1.02$; Fig.~\ref{f02}B), high monitoring costs and low exit incentives yield a stationary state consisting of cooperators and defectors only ($C{+}D$). As the exit incentive $\epsilon$ increases and/or the monitoring cost $c$ decreases, all networks undergo a transition through an intermediate $C{+}D{+}DE$ regime and, at sufficiently low monitoring costs, reach a cooperative–exit mixture $C{+}CE{+}DE$. An oscillatory coexistence state $O$ emerges across all networks at low $c$ and high $\epsilon$ (visualized below). Compared with the regular lattice, the regular small-world network and the random regular graph exhibit an additional low-$c$, low-$\epsilon$ wedge ($C{+}D{+}CE$), which is most pronounced in the random regular graph.

For the strong dilemma case ($b=1.5$; Fig.~\ref{f02}C), cooperative coexistence is less constrained than in well-mixed populations and persists even when $\epsilon<c$. Across all three regular networks, high monitoring costs lead to a defector-dominated stationary state ($D$). As $\epsilon$ increases and/or $c$ decreases, the system transitions sequentially from $D$ to $C{+}D{+}CE$, followed by full four-strategy coexistence ($C{+}D{+}CE{+}DE$), and finally to a cooperative–exit regime without defectors ($C{+}CE{+}DE$). An exception arises in the random regular graph, where much of the low-cost cooperative–exit region is replaced by the oscillatory state $O$.

To quantify how exit incentives reshape relative strategy success, we vary $\epsilon$ at a fixed monitoring cost $c=0.4$ (Fig.~\ref{f03}). For the weak dilemma ($b=1.02$; Fig.~\ref{f03}A), low exit incentives produce a $C{+}D$ regime in which defectors are more abundant than cooperators and exit strategies remain rare. As $\epsilon$ increases, defective exiters ($DE$) rise in frequency while defectors decline and cooperators increase, ultimately giving way to the oscillatory state $O$. This exit-incentive–dependent reordering of strategy frequencies is observed across all network topologies, although the precise transition points differ.

For the strong dilemma ($b=1.5$; Fig.~\ref{f03}B), low exit incentives lead to complete defection across all networks. As $\epsilon$ increases, the system first enters a $C{+}D{+}CE$ regime, in which cooperators and cooperative exiters ($CE$) expand at the expense of defectors. Upon entering the $C{+}D{+}CE{+}DE$ regime, defective exiters ($DE$) also emerge, accompanied by further reductions in defection. At high exit incentives, the lattice and small-world networks converge to a cooperative–exit regime ($C{+}CE{+}DE$), whereas the random regular graph instead transitions to the oscillatory state $O$, in which all four strategies persist and defectors remain at low average frequency.

Microscopic inspection reveals that conditional exiters act as key mediators of exit-incentive–dependent coexistence by inducing cyclic dominance at the local level. Fig.~\ref{f04} illustrates how local interactions generate distinct coexistence routes for cooperation by showing the spatiotemporal dynamics of strategy distributions on the regular lattice for dilemma strength $b=1.5$ and monitoring cost $c=0.4$, as the exit incentive $\epsilon$ increases.

When the exit payoff is low ($\epsilon=0.1$; top row of Fig.~\ref{f04}), monitoring costs dominate. Defective exiters ($DE$) are rapidly eliminated by defectors ($D$), while conditional exiters ($CE$) transiently shield cooperators ($C$) by forming protective clusters. However, $CE$ cannot resist invasion by $D$, and the erosion of these clusters ultimately leads to the collapse of cooperation.

At intermediate exit incentives ($\epsilon=0.4$; second row), the exit payoff equals the monitoring cost, qualitatively altering local competition. Although $DE$ remain unviable, $CE$ successfully invade $D$, while cost-free $C$ invade $CE$. This establishes a three-strategy cyclic dominance loop, $C \to D \to CE \to C$, in which domains are continually created and destroyed, allowing cooperation to persist dynamically.

As the exit incentive increases further ($\epsilon=0.5$; third row), $DE$ become viable and can withstand competition from $D$. Defective exiters invade defector domains and, in turn, give rise to cooperative exiters, extending the interaction chain. The system self-organizes into a four-strategy cyclic dominance loop, $C \to D \to DE \to CE \to C$, producing high diversity and persistent spatiotemporal fluctuations.

For substantial exit payoffs ($\epsilon=0.9$; bottom row), reputation-based strategies gain a decisive advantage. Unconditional defectors are rapidly suppressed by $DE$ and disappear from the population. The dynamics then reduce to a three-strategy loop involving $C$, $CE$, and $DE$: $C \to DE \to CE \to C$, through which cooperation is maintained by continual local cycling.

These exit-incentive–dependent coexistence routes and associated dominance relations are qualitatively consistent across regular lattices, small-world networks, and random regular graphs (Figs.~\ref{f02}–\ref{f03}). Accordingly, the mediating role of conditional exiters is not restricted to lattices but extends to other regular network structures. Consistent with their shorter characteristic path lengths~\cite{watts1998collective}, random regular graphs exhibit the broadest parameter regions supporting oscillatory coexistence.

\begin{figure}[t]
	\centering
	\includegraphics[scale=0.55]{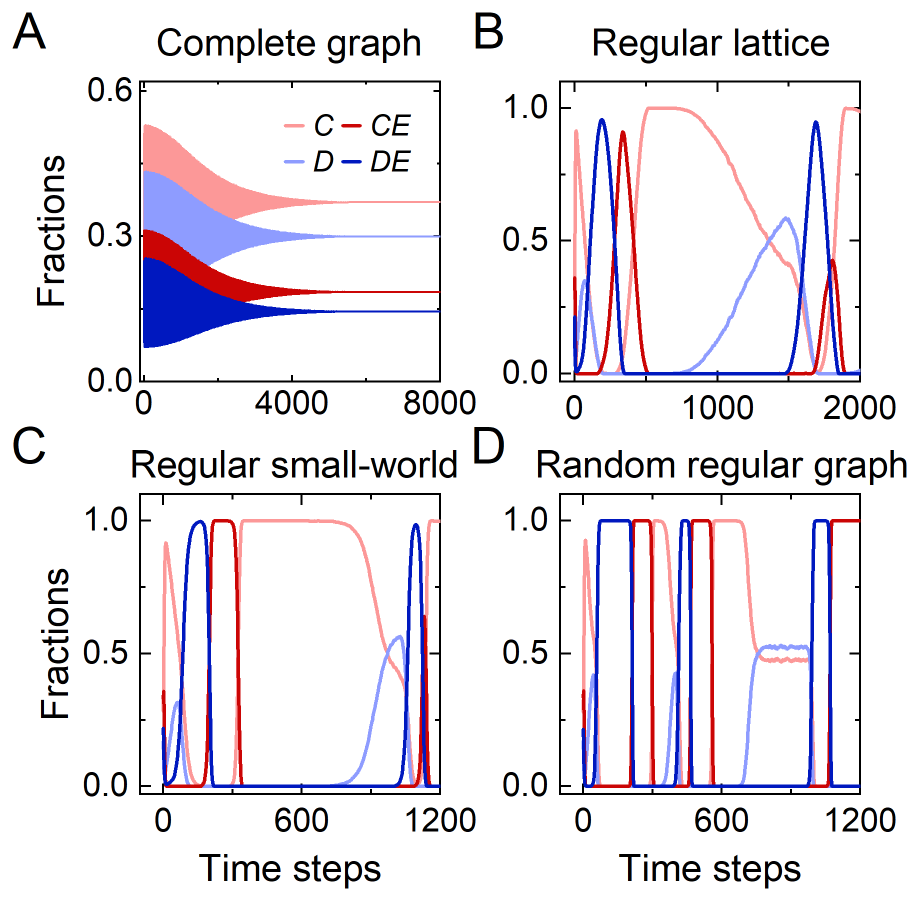}
	\caption{\textbf{Structured populations can generate persistent oscillatory dynamics.} Time series of strategy fractions for (A) a complete graph (well-mixed population), (B) a regular lattice, (C) a regular small-world network, and (D) a random regular graph. To illustrate representative dynamics across population structures, the dilemma strength is set to $b=1.5$ for the complete graph and to $b=1.02$ for the structured populations. The monitoring cost and exit incentive are fixed at $c=0.4$ and $\epsilon=0.9$, respectively. Time axes differ across panels for visualization purposes. 
	}
	\label{f05}
\end{figure} 

Fig.~\ref{f05} contrasts the temporal evolution of strategy frequencies in well-mixed and structured populations under parameter combinations corresponding to the oscillatory regime. In the complete graph (Fig.~\ref{f05}A), the population rapidly converges to a stationary mixed state, with all strategy frequencies approaching constant values. No sustained oscillations are observed, consistent with replicator dynamics in well-mixed populations.

In contrast, all three structured populations---the regular lattice, small-world network, and random regular graph---also exhibit persistent oscillations in strategy frequencies, in addition to the locally cyclic dominance illustrated in Fig.~\ref{f04}. As shown in Fig.~\ref{f05}B–D, after transient dynamics, the abundances of cooperators, defectors, conditional cooperative exiters, and conditional defective exiters continue to fluctuate with finite macroscopic amplitude and follow a four-strategy cyclic dominance loop, $C \rightarrow D \rightarrow DE \rightarrow CE \rightarrow C$, indicating long-lived nonstationary coexistence. Although oscillations arise across all structured networks, their amplitude and temporal regularity depend on network topology. These sustained oscillations correspond to the oscillatory coexistence state $O$ identified in the phase diagrams (Figs.~\ref{f02}–\ref{f03}), confirming that $O$ represents genuinely nonstationary dynamics unique to structured populations in the proposed reputation-based voluntary Prisoner’s Dilemma.

\section{Discussions}
\label{discussions}
We developed a reputation-based voluntary Prisoner’s Dilemma game to examine how information-conditioned exit decisions shape the evolutionary dynamics of cooperation. In this game, players incur a monitoring cost to inspect opponents’ strategies as reputation signals, enabling conditional exit from interactions. We show that in well-mixed populations, reputation-based exit sustains stable cooperative coexistence only when exit incentives exceed monitoring costs. By contrast, population structure relaxes this requirement and enables multiple exit-incentive–dependent pathways to cooperation. Across regular networks, exit incentives interact with spatial structure to produce qualitatively distinct dynamics, ranging from locally mediated cyclic dominance to persistent population-level oscillations.

Placed within the literature on voluntary participation, our results broaden the scope of this mechanism by introducing reputation-conditioned participation and revealing richer evolutionary outcomes. Classical voluntary participation models typically treat opting out as an unconditional, information-free decision or embed exit choices within repeated-interaction settings. By contrast, we study reputation-conditioned voluntary participation in a one-shot game, without restricting attention to anonymity or small-group repetition. Existing voluntary participation models sustain cooperation primarily through rock–paper–scissors–type cyclic dominance~\cite{hauert2002replicator,hauert2007via,szabo2002phase} or through the success of particular strategies in repeated games~\cite{izquierdo2010option,izquierdo2014leave}. We show that conditioning exit on reputation generates exit-incentive–dependent coexistence pathways, including locally driven cyclic dominance and persistent nonstationary oscillations. Although derived here in a reputation-based voluntary Prisoner’s Dilemma game, the underlying participation mechanism is general and can be naturally extended to multiplayer social dilemmas, such as public goods games, warranting further investigation.

Beyond these theoretical insights, our findings offer insights relevant to distributed multi-agent systems in which participation decisions depend on locally available information. In peer-to-peer networks, decentralized marketplaces, and online collaboration platforms, agents routinely assess potential partners before deciding whether to engage or withdraw. Our results indicate that the exit incentive functions as a key control parameter that offsets the costs associated with acquiring and using reputation information, rather than acting as a passive participation subsidy. Its magnitude determines the resulting coexistence regime, the prevalence of cooperation, and whether system-level dynamics are stationary or oscillatory. Accordingly, system designers should jointly calibrate exit incentives and interaction topology to promote robust cooperation while avoiding parameter regimes that induce large-amplitude, system-wide oscillations.

Finally, our reputation-based voluntary Prisoner’s Dilemma game points to several promising directions for future research. First, we adopted a first-order information assumption, in which defectors are uniformly identified as undesirable. In contrast, reputation in indirect reciprocity often depends on higher-order information, incorporating opponents’ past interactions and social context. Incorporating explicit and evolving reputation dynamics, including subjective or noisy evaluations, would allow future work to examine how informational complexity reshapes cooperative coexistence pathways. Second, our analysis focused on static regular networks, whereas in many real systems the same information that enables conditional exit may also drive structural adaptation. Exploring the coevolution of network structure and reputation-based voluntary participation therefore represents a natural extension of this work and may uncover additional evolutionary pathways to cooperation.

\section{Article information}
\paragraph*{Acknowledgments} 
We acknowledge support from (i) the National Science Fund for Distinguished Young Scholars (grant no. 62025602), the National Natural Science Foundation of China (grant no. U22B2036) and the XPLORER PRIZE 2021 of the Tencent Foundation; (ii) the National Natural Science Foundation of China (grant nos. 12271471 and 11931015), Major Program of National Fund of Philosophy and Social Science of China (grants Nos. 22\&ZD158 and 22VRCO49) to L.\, S. and (iii) the JSPS KAKENHI (grant no. JP 23H03499). M.P. was supported by the Slovenian Research and Innovation Agency (grant no. P1-0403).

\paragraph*{Author contributions.} 
C.\,S. conceived research. C.\,S., Z.\, S., and X.\,W. performed theoretical analysis and evolutionary simulations. All co-authors discussed the results and wrote the manuscript.
\paragraph*{Conflict of interest.} Authors declare no conflict of interest.

\bibliographystyle{IEEEtran}
\bibliography{biblio}

\end{document}